\documentclass{PoS}

\title{Parameter degeneracy and hierarchy sensitivity of NO$\nu$A in presence of sterile neutrino}

\ShortTitle{Degeneracies with sterile neutrinos in NO$\nu$A}

\author{\speaker{Monojit Ghosh}\\
        Department of Physics, Tokyo Metropolitan University, Hachioji, Tokyo 192-0397, Japan\\
        E-mail: \email{mghosh@phys.se.tmu.ac.jp}}

\author{Shivani Gupta\\
        Center of Excellence for Particle Physics at the Terascale (CoEPP), University of Adelaide, Adelaide SA 5005, Australia\\
        E-mail: \email{shivani.gupta@adelaide.edu.au}}
        
\author{Zachary M. Matthews\\
        Center of Excellence for Particle Physics at the Terascale (CoEPP), University of Adelaide, Adelaide SA 5005, Australia\\
        E-mail: \email{zachary.matthews@adelaide.edu.au}}
        
\author{Pankaj Sharma\\
        Center of Excellence for Particle Physics at the Terascale (CoEPP), University of Adelaide, Adelaide SA 5005, Australia\\
        E-mail: \email{pankaj.sharma@adelaide.edu.au}}
        
\author{Anthony G. Williams\\
        Center of Excellence for Particle Physics at the Terascale (CoEPP), University of Adelaide, Adelaide SA 5005, Australia\\
        E-mail: \email{anthony.williams@adelaide.edu.au}}        

\abstract{The first hint of neutrino mass hierarchy is expected to come from the NO$\nu$A experiment in Fermilab as the
present best-fit parameter space i.e., normal hierarchy and $\delta_{CP}=-90^\circ$ is the favourable parameter space for NO$\nu$A where there is no degeneracy. But this situation may change if
the standard three flavour framework is not complete and there is existence of new physics. In this work we consider the presence of an extra light sterile neutrino at the eV scale and study the
new degeneracies which are absent in the standard three flavour framework. We also study the effect of these new degeneracies on the hierarchy measurement of NO$\nu$A.}

\FullConference{The 19th International Workshop on Neutrinos from Accelerators-NUFACT2017\\
		25-30 September, 2017\\
		Uppsala University, Uppsala, Sweden}

\begin{document}

\section{Introduction}

In the standard three flavour framework, neutrino oscillation in which neutrinos change their flavour is described by six parameters: three mixing angles: $\theta_{12}$, $\theta_{13}$, $\theta_{23}$,
two mass squared differences: $\Delta_{21}$ ($m_2^2 - m_1^2$), and $\Delta_{31}$ ($m_3^2 - m_1^2$) and one phase $\delta_{13}$. Among them one of the major unknown is the sign of $\Delta_{31}$ or
the neutrino mass hierarchy. It can be either normal i.e., $\Delta_{31} > 0$ (NH) or inverted i.e., $\Delta_{31} < 0$ (IH). It is well known that if Nature choose the favourable parameter space 
where there is no degeneracy, then NO$\nu$A \cite{Adamson:2017gxd} can determine neutrino mass hierarchy at more that $2 \sigma$ C.L. 
Fortunately the current best fit parameter space i.e., NH with $\delta_{13}=-90^\circ$ \cite{Forero:2014bxa,Esteban:2016qun,Capozzi:2013csa}
is indeed the favourable parameter space for NO$\nu$A and thus it is expected that the first hint of neutrino mass hierarchy will come from the NO$\nu$A experiment. But the situation can be different
if there exists new physics. In presence of new physics there can be additional degeneracies which can spoil the hierarchy sensitivity of NO$\nu$A even for the favourable parameter space.
In this work we consider the existence of an extra light sterile neutrino at the eV scale \cite{Abazajian:2012ys} i.e. the 3+1 scenario. 
In this present work our aim is to identify the new degeneracies and study their
effect in the determination of hierarchy in NO$\nu$A.

\section{Oscillation parameters in 3+1 scheme}
In presence of one extra light sterile neutrino, we parametrize the PMNS matrix as
\begin{table}[h]
	\centering
	\begin{tabular}{|c|c|c|}
		\hline
		$4\nu$ Parameters & True Value & Test Value Range\\
		\hline
		$\sin^2\theta_{12}$ & $0.304$ & $\mathrm{N/A}$\\
		$\sin^22\theta_{13}$ & $0.085$ & $\mathrm{N/A}$\\
		$\theta_{23}^{\mathrm{LO}}$ & $40^\circ$ & $(40^\circ,50^\circ)$\\
		$\theta_{23}^{\mathrm{HO}}$ & $50^\circ$ & $(40^\circ,50^\circ)$\\
		$\sin^2\theta_{14}$ & $0.025$ & $\mathrm{N/A}$\\
		$\sin^2\theta_{24}$ & $0.025$ & $\mathrm{N/A}$\\
		$\theta_{34}$ & $0^\circ$ & $\mathrm{N/A}$\\
		$\delta_{13}$ & $-90^\circ$ & $(-180^\circ,180^\circ)$\\
		$\delta_{14}$ &  $-90^\circ,0^\circ,90^\circ$ & $(-180^\circ,180^\circ)$\\
		$\delta_{34}$ & $0^\circ$ & $\mathrm{N/A}$\\
		$\Delta_{21}$ & $7.5\times10^{-5}\mathrm{eV}^2$ & $\mathrm{N/A}$\\
		$\Delta_{31}$ & $2.475\times10^{-3}\mathrm{eV}^2$ & $(2.2,2.6)\times10^{-3}\mathrm{eV}^2$\\
		$\Delta_{41}$ & $1\mathrm{eV}^2$ & $\mathrm{N/A}$\\
		\hline
	\end{tabular}
	\caption{\label{tab:i} Expanded $4\nu$ parameter true values and test marginalisation ranges, parameters with N/A are not marginalised over.
	\label{SterParam}}
\end{table}
\begin{equation}
U_{\mathrm{PMNS}}^{4\nu}=
U(\theta_{34},\delta_{34})
U(\theta_{24},0)
U(\theta_{14},\delta_{14})
U_{\mathrm{PMNS}}^{3\nu}\,.
\end{equation}
where
\begin{equation}
U_{\mathrm{PMNS}}^{3\nu}
=
U(\theta_{23},0)
U(\theta_{13},\delta_{13})
U(\theta_{12},0)\,.
\end{equation}where $U(\theta_{ij},\delta_{ij})$ contains a corresponding $2\times2$ mixing matrix:
\begin{equation}
U^{2\times 2}(\theta_{ij},\delta_{ij})
=
\left(
\begin{array}{c c}
\mathrm{c}_{ij} & \mathrm{s}_{ij}e^{i\delta_{ij}}\\
-\mathrm{s}_{ij}e^{i\delta_{ij}} & \mathrm{c}_{ij}
\end{array}
\right)
\end{equation}embedded in an $n\times n$ array in the $i,j$ sub-block.
Thus in this case the neutrino oscillation parameter space is increased by three more mixing angles: $\theta_{14}$, $\theta_{24}$ and $\theta_{34}$, two more Dirac type
CP phases i.e., $\delta_{14}$ and $\delta_{34}$ and one more mass squared difference: $\Delta_{41}$ ($m_4^2 - m_1^2$). 
In 3+1 case, the appearance channel expression in vacuum is given by \cite{Klop:2014ima}
\begin{eqnarray} \nonumber
\label{eq:Pme_atm}
 & P_{\mu e} &\!\! \simeq\,  4 s_{23}^2 s^2_{13}  \sin^2{\Delta} +  8 s_{13} s_{12} c_{12} s_{23} c_{23} (\alpha \Delta)\sin \Delta \cos({\Delta \pm \delta_{13}}) +
   4 s_{14} s_{24} s_{13} s_{23} \sin\Delta \sin (\Delta \pm \delta_{13} \mp  \delta_{14})
\end{eqnarray}
where $\Delta \equiv  \Delta_{31}L/4E$, $\alpha \equiv \Delta_{21}/ \Delta_{31}$ with $L$ being the baseline and $E$ is the energy.
For our present work we list our choice of parameters in Table \ref{tab:i} \cite{Kopp:2013vaa}.

\begin{figure*}[h]\centering
\includegraphics[scale=0.75]{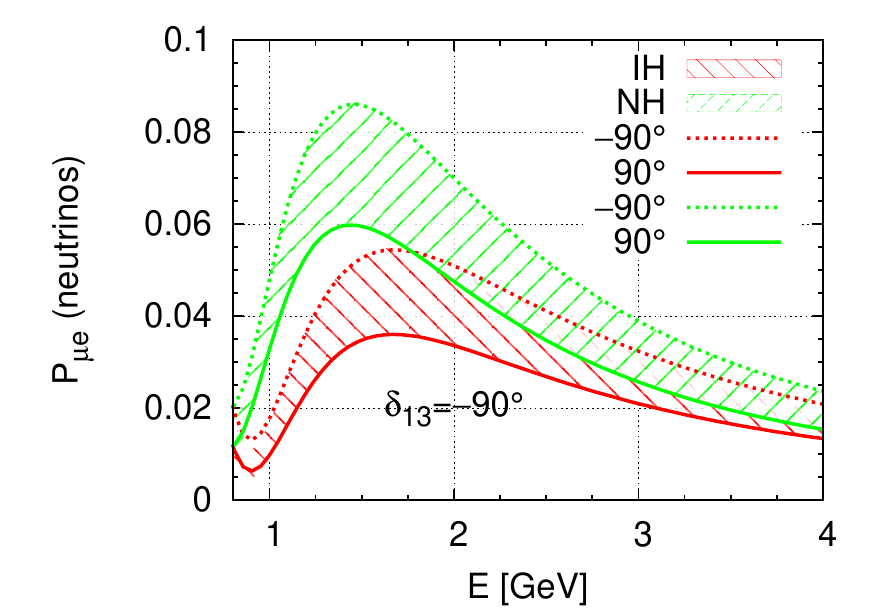}
\includegraphics[scale=0.75]{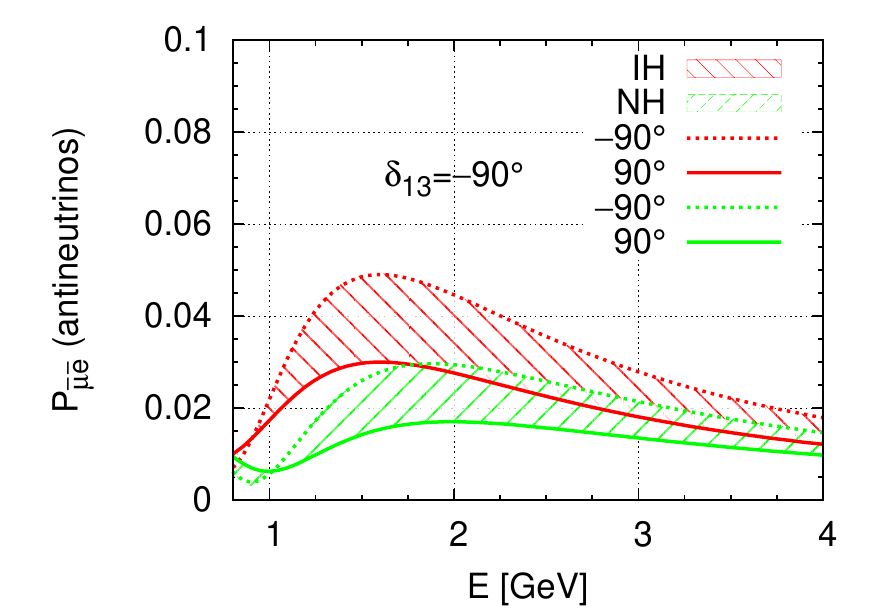}\\
\includegraphics[scale=0.75]{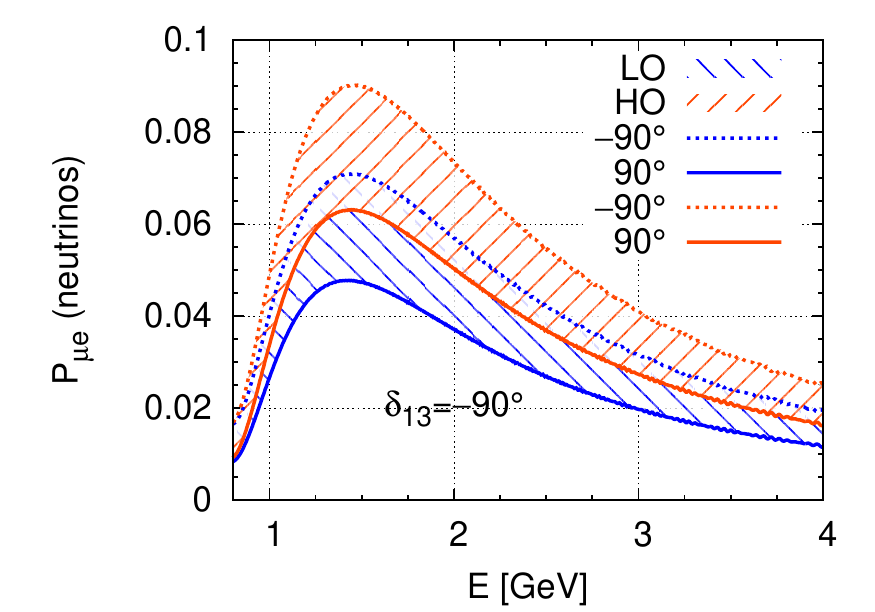}
\includegraphics[scale=0.75]{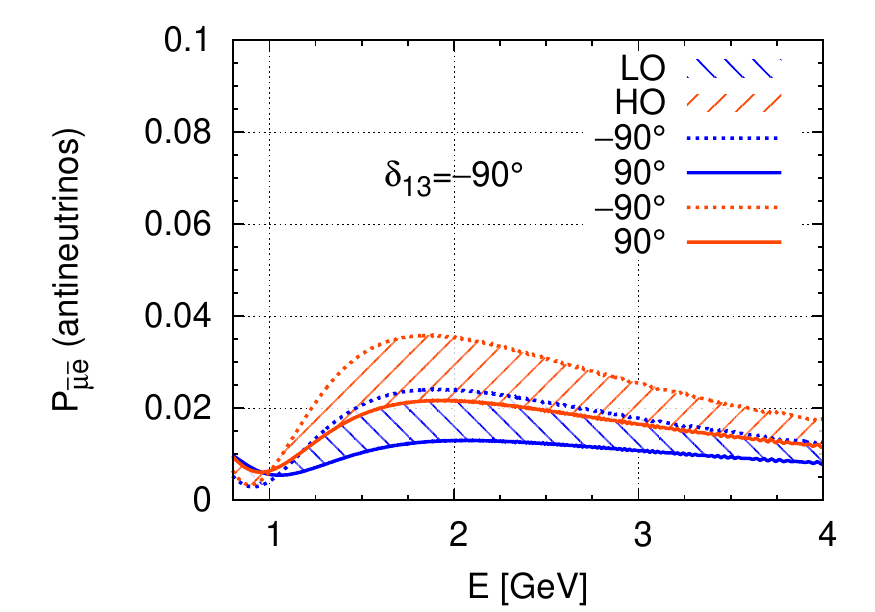}
\caption{$\nu_\mu \rightarrow\nu_e$ oscillation probability bands for $\delta_{13} = -90^\circ$. Left panels are for neutrinos and right panels are for antineutrinos. 
The upper panel shows the hierarchy-$\delta_{14}$ degeneracy and the lower panels shows the octant-$\delta_{14}$ degeneracy.}
\label{fig:prob}
\end{figure*}
\begin{figure*}
     \includegraphics[width=0.45\textwidth]{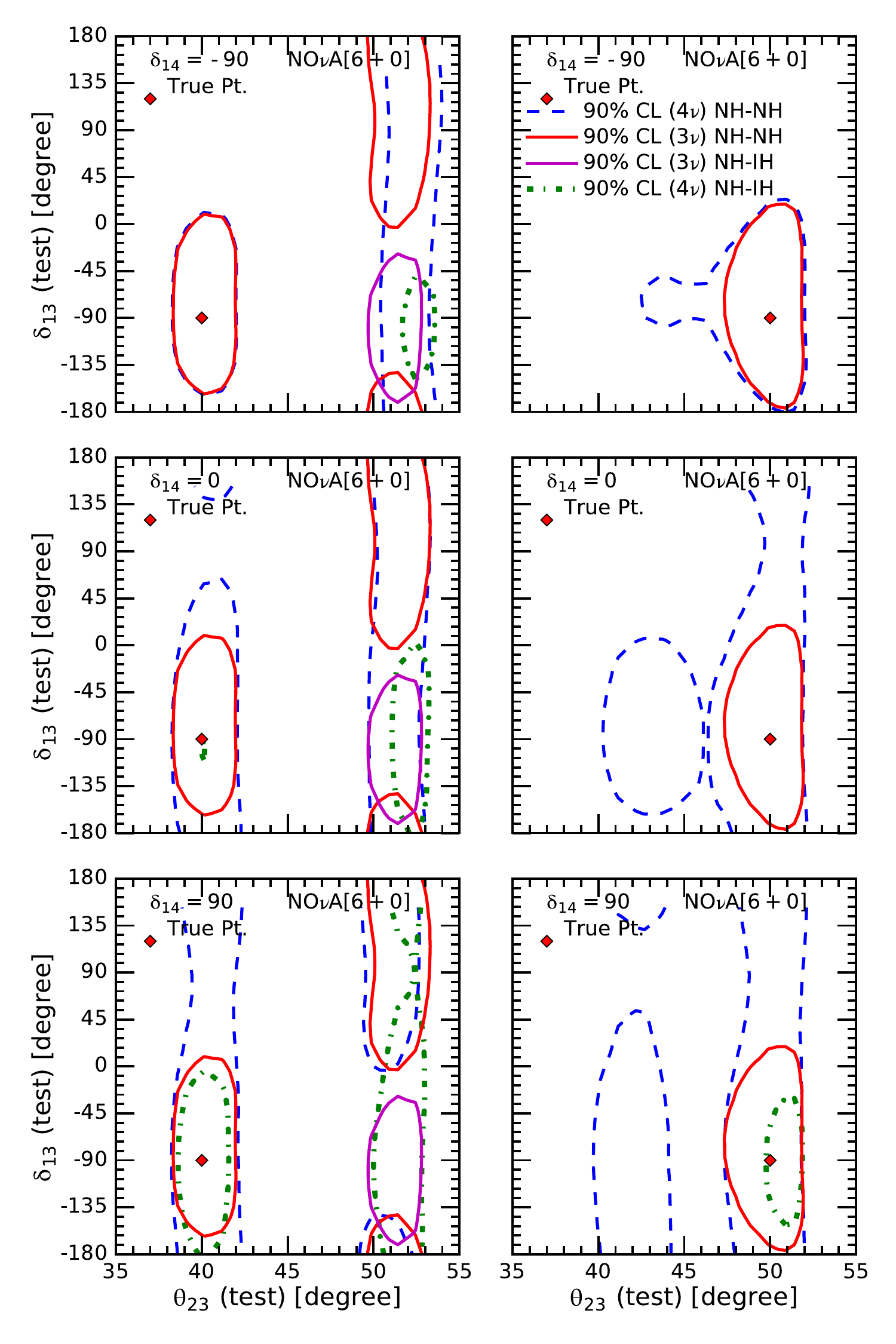}
     \includegraphics[width=0.45\textwidth]{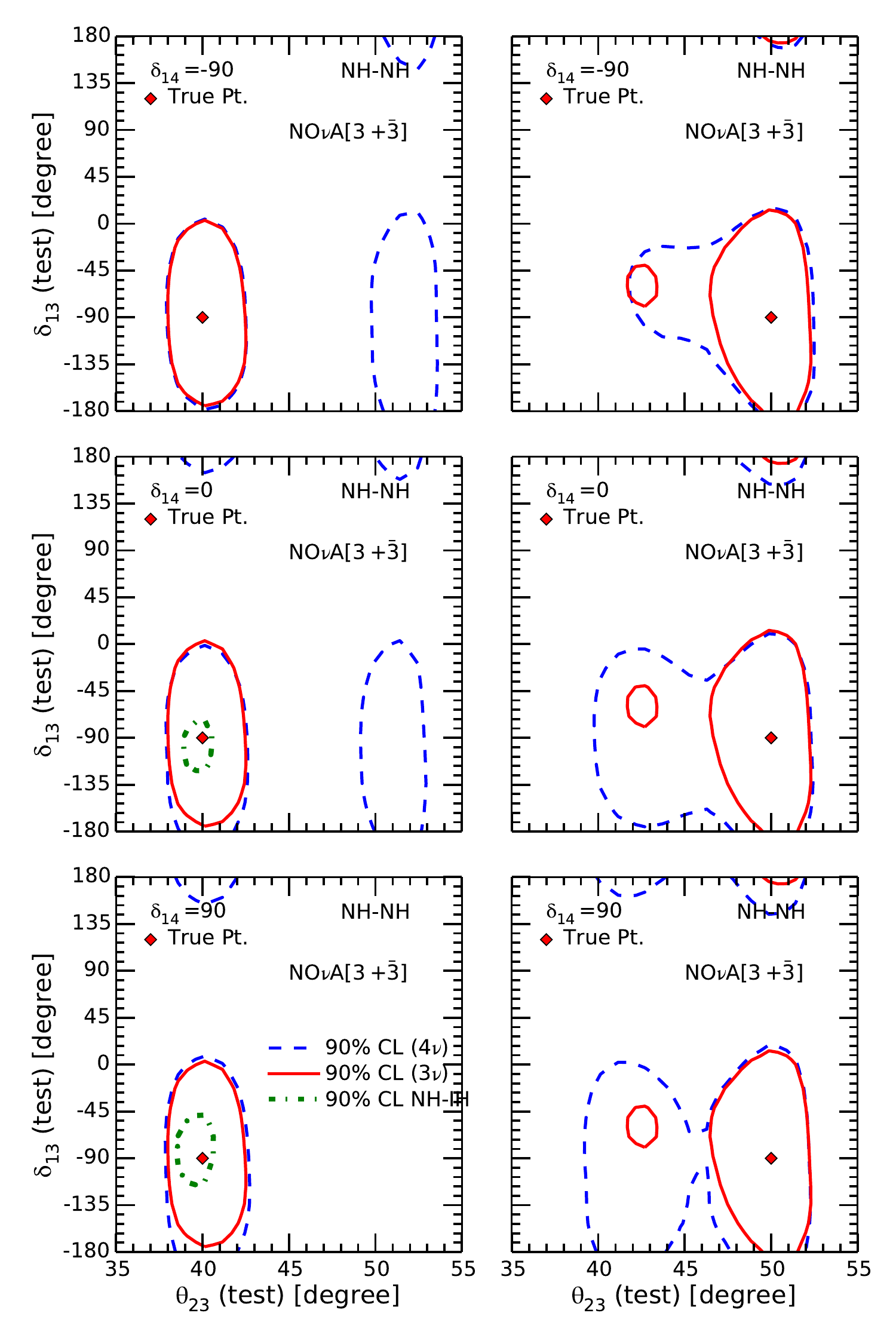}
     \caption{Contour plots in the $\theta_{23}({\rm test})$ vs $\delta_{13}({\rm test})$ plane for two different true values of $\theta_{23}= 40^\circ$ (first and third column) and $50^\circ$ (second and fourth column) for NO$\nu$A $(6+\bar0)$ (first and second column) and ($3+\bar 3$) (third and fourth column). The first, second and third rows are for $\delta_{14}=-90^\circ$ , $0^\circ$ and $90^\circ$ respectively. The true value for the $\delta_{13}$ is taken to be $-90^\circ$. The true hierarchy is NH. We marginalize over the test values of $\delta_{14}$. Also shown is the contours for the $3\nu$ flavor scenario.}
     \label{fig:hierarchy_degen}
\end{figure*}
\begin{figure}\centering
     \includegraphics[width=0.45\textwidth]{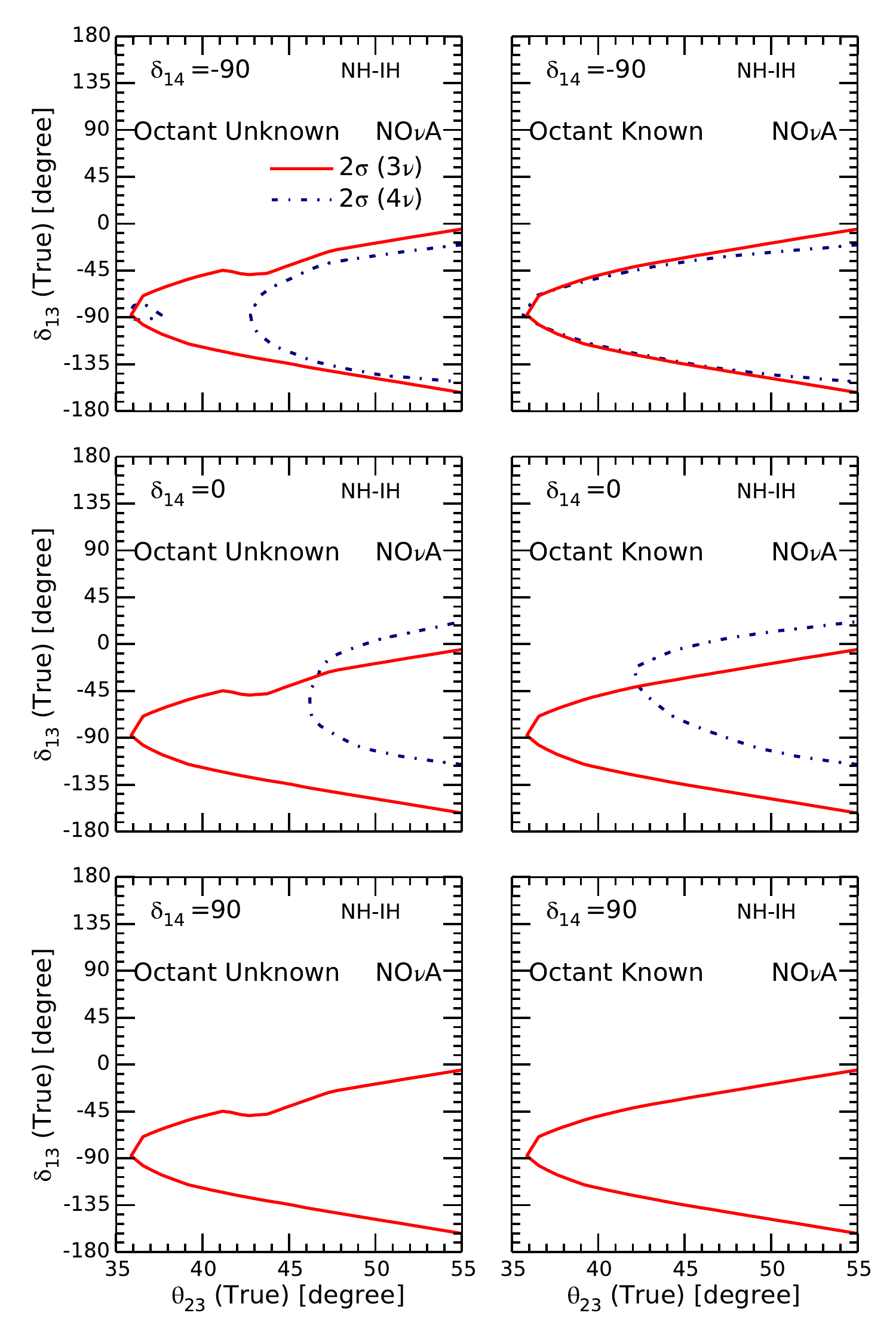}
     \caption{Contour plots at $2\sigma$ C.L. in the $\theta_{23}({\rm true})$ vs $\delta_{13}({\rm true})$ plane for Octant Unknown (left panel) and Octant Known (right panel) scenarios for NO$\nu$A ($3+\bar 3$). The first, second and third rows are for $\delta_{14}=-90^\circ$, $0^\circ$ and $90^\circ$ respectively. The true and test hierarchies are chosen to be normal (NH) and inverted hierarchy (IH) respectively. Also shown contours for the $3\nu$ flavor scenario.}
     \label{fig:hierarchy_degen_1}
\end{figure}
In the table, LO implies the lower octant of $\theta_{23}$ and HO implies higher octant of $\theta_{23}$. We have generated all our results with the GLoBES software \cite{Huber:2004ka}.

\section{Degeneracy at the probability level}
In fig. \ref{fig:prob}, we have plotted the appearance probability vs energy for $\delta_{13} = -90^\circ$ and the bands are due to variation of $\delta_{14}$. The upper panels show the
hierarchy-$\delta_{14}$ degeneracy and the lower panels depict octant-$\delta_{14}$ degeneracy. 
From the upper panels we see that we have degeneracies in \{NH, $\delta_{14}=90^\circ$\} with \{IH, $\delta_{14}=-90^\circ$\} for neutrinos and 
\{NH, $\delta_{14}=-90^\circ$\} with \{IH, $\delta_{14}=90^\circ$\} for antineutrinos. Thus we understand that this degeneracy can be removed with a balanced run of neutrinos and antineutrinos.
From the lower panels we see that there is degeneracies in \{LO, $\delta_{14}=-90^\circ$ \} with \{HO, $\delta_{14}=90^\circ$\} for both neutrinos and antineutrinos. Thus it is clear that this
degeneracy is unremovable. It was shown in Ref. \cite{Agarwalla:2016xlg} that due to this degeneracy, the octant determination of the long-baseline experiments is highly compromised.

\section{Degeneracies at the event level}
To Show the degeneracies at the event level, in Fig. \ref{fig:hierarchy_degen} we have given the contour plots in the $\theta_{23}$ (test) - $\delta_{13}$ (test) plane. The true point
is represented by the red diamond. In these panels,
red and purple contours correspond to the right hierarchy and wrong hierarchy solutions respectively for the three generation case and the blue and green contours correspond to right hierarchy 
and wrong hierarchy solutions respectively for the 3+1 case. By comparing the pure neutrino results of NO$\nu$A labeled as NO$\nu$A (6+0) and mixed neutrino-antineutrino results labeled
as NO$\nu$A (3+3) we notice that for three generation case, the all the degenerate solutions are almost gone when antineutrino data is considered. 
But for the 3+1 case, we notice that the wrong hierarchy solutions are almost gone but
the wrong octant solutions does not get removed.

Thus from the above discussion we understand that even for NH and $\delta_{13} = -90^\circ$, where there is almost no degeneracy in the three generation case, there exists degenerate solutions 
when there is an extra light sterile neutrino.

\section{Results for hierarchy sensitivity}
To study the effect of these degeneracies on the hierarchy measurement in Fig. \ref{fig:hierarchy_degen_1} we have plotted the hierarchy $\chi^2$ in the true $\theta_{23}$ -true $\delta_{13}$ plane.
From the figure we see that NO$\nu$A has good hierarchy sensitivity for $\delta_{13}=-90^\circ$ for the generation case. 
But for the 3+1 case, the sensitivity depends on the true value of $\delta_{14}$. For $\delta_{14}=-90^\circ$ we see that the hierarchy sensitivity is lost for $\theta_{23} < 43^\circ$ if the
octant is unknown. However if the octant is known then the sensitivity coincides with the three generation case. For $\delta_{14} = 0^\circ$, we note that the hierarchy sensitivity is lost
if $\theta_{23}$ is less than $46^\circ$ for both the cases. However the most remarkable result is obtained if $\delta_{14}$ is $90^\circ$. In this case we see that there is a complete loss of
hierarchy sensitivity at $2 \sigma$ for all true values of $\theta_{23}$.

\section{Summary}

In this work we have studied the parameter degeneracy in neutrino oscillation in the presence of a light sterile neutrino in the eV scale for NO$\nu$A. In our work we have identified new degeneracies
which are absent in the standard three generation case. Because of these there are unsolved degenerate region in the 3+1 case. We also showed that the hierarchy sensitivity depends on the
true values of $\theta_{14}$. If the observed hierarchy sensitivity of NO$\nu$A is less than the expected then this can be a hint of existence of sterile neutrinos. For more detail we refer to
\cite{Ghosh:2017atj} on which this article is based upon.

\section*{Acknowledgements}
The work of MG is partly supported by the ``Grant-in-Aid for Scientific Research of the Ministry of Education, Science and Culture, Japan", under Grant No. 25105009. 
SG, ZM, PS and AGW acknowledge the support by the University of Adelaide and the Australian Research Council through the ARC Centre of Excellence for 
Particle Physics at the Terascale (CoEPP) (grant no. CE110001004).


\begin{thebibliography}{99}

%\cite{Forero:2014bxa}
\bibitem{Forero:2014bxa} 
  D.~V.~Forero, M.~Tortola and J.~W.~F.~Valle,
  %``Neutrino oscillations refitted,''
  Phys.\ Rev.\ D {\bf 90}, no. 9, 093006 (2014)
%  doi:10.1103/PhysRevD.90.093006
  [arXiv:1405.7540 [hep-ph]].
  %%CITATION = doi:10.1103/PhysRevD.90.093006;%%
  %502 citations counted in INSPIRE as of 13 Dec 2017
  
  %\cite{Esteban:2016qun}
\bibitem{Esteban:2016qun} 
  I.~Esteban, M.~C.~Gonzalez-Garcia, M.~Maltoni, I.~Martinez-Soler and T.~Schwetz,
  %``Updated fit to three neutrino mixing: exploring the accelerator-reactor complementarity,''
  JHEP {\bf 1701}, 087 (2017)
%  doi:10.1007/JHEP01(2017)087
  [arXiv:1611.01514 [hep-ph]].
  %%CITATION = doi:10.1007/JHEP01(2017)087;%%
  %183 citations counted in INSPIRE as of 13 Dec 2017

  %\cite{Capozzi:2013csa}
\bibitem{Capozzi:2013csa} 
  F.~Capozzi, G.~L.~Fogli, E.~Lisi, A.~Marrone, D.~Montanino and A.~Palazzo,
  %``Status of three-neutrino oscillation parameters, circa 2013,''
  Phys.\ Rev.\ D {\bf 89}, 093018 (2014)
%  doi:10.1103/PhysRevD.89.093018
  [arXiv:1312.2878 [hep-ph]].
  %%CITATION = doi:10.1103/PhysRevD.89.093018;%%
  %473 citations counted in INSPIRE as of 13 Dec 2017


%\cite{Adamson:2017gxd}
\bibitem{Adamson:2017gxd} 
  P.~Adamson {\it et al.} [NOvA Collaboration],
  %``Constraints on Oscillation Parameters from $\nu_e$ Appearance and $\nu_\mu$ Disappearance in NOvA,''
  Phys.\ Rev.\ Lett.\  {\bf 118}, no. 23, 231801 (2017)
%  doi:10.1103/PhysRevLett.118.231801
  [arXiv:1703.03328 [hep-ex]].
  %%CITATION = doi:10.1103/PhysRevLett.118.231801;%%
  %33 citations counted in INSPIRE as of 13 Dec 2017


%\cite{Abazajian:2012ys}
\bibitem{Abazajian:2012ys} 
  K.~N.~Abazajian {\it et al.},
  %``Light Sterile Neutrinos: A White Paper,''
  arXiv:1204.5379 [hep-ph].
  %%CITATION = ARXIV:1204.5379;%%
  %596 citations counted in INSPIRE as of 13 Dec 2017
  
  %\cite{Klop:2014ima}
\bibitem{Klop:2014ima} 
  N.~Klop and A.~Palazzo,
  %``Imprints of CP violation induced by sterile neutrinos in T2K data,''
  Phys.\ Rev.\ D {\bf 91}, no. 7, 073017 (2015)
%  doi:10.1103/PhysRevD.91.073017
  [arXiv:1412.7524 [hep-ph]].
  %%CITATION = doi:10.1103/PhysRevD.91.073017;%%
  %39 citations counted in INSPIRE as of 13 Dec 2017
  
  %\cite{Kopp:2013vaa}
\bibitem{Kopp:2013vaa} 
  J.~Kopp, P.~A.~N.~Machado, M.~Maltoni and T.~Schwetz,
  %``Sterile Neutrino Oscillations: The Global Picture,''
  JHEP {\bf 1305}, 050 (2013)
  doi:10.1007/JHEP05(2013)050
  [arXiv:1303.3011 [hep-ph]].
  %%CITATION = doi:10.1007/JHEP05(2013)050;%%
  %364 citations counted in INSPIRE as of 13 Dec 2017

  
  %\cite{Huber:2004ka}
\bibitem{Huber:2004ka} 
  P.~Huber, M.~Lindner and W.~Winter,
  %``Simulation of long-baseline neutrino oscillation experiments with GLoBES (General Long Baseline Experiment Simulator),''
  Comput.\ Phys.\ Commun.\  {\bf 167}, 195 (2005)
%  doi:10.1016/j.cpc.2005.01.003
  [hep-ph/0407333].
  %%CITATION = doi:10.1016/j.cpc.2005.01.003;%%
  %443 citations counted in INSPIRE as of 13 Dec 2017
  
  %\cite{Agarwalla:2016xlg}
\bibitem{Agarwalla:2016xlg} 
  S.~K.~Agarwalla, S.~S.~Chatterjee and A.~Palazzo,
  %``Octant of $\theta_{23}$ in danger with a light sterile neutrino,''
  Phys.\ Rev.\ Lett.\  {\bf 118}, no. 3, 031804 (2017)
%  doi:10.1103/PhysRevLett.118.031804
  [arXiv:1605.04299 [hep-ph]].
  %%CITATION = doi:10.1103/PhysRevLett.118.031804;%%
  %23 citations counted in INSPIRE as of 13 Dec 2017
  
  %\cite{Ghosh:2017atj}
\bibitem{Ghosh:2017atj} 
  M.~Ghosh, S.~Gupta, Z.~M.~Matthews, P.~Sharma and A.~G.~Williams,
  %``Study of parameter degeneracy and hierarchy sensitivity of NO$\nu$A in presence of sterile neutrino,''
  Phys.\ Rev.\ D {\bf 96}, no. 7, 075018 (2017)
%  doi:10.1103/PhysRevD.96.075018
  [arXiv:1704.04771 [hep-ph]].
  %%CITATION = doi:10.1103/PhysRevD.96.075018;%%
  %4 citations counted in INSPIRE as of 13 Dec 2017




\end{thebibliography}
\end{document}